\documentclass[prl,twocolumn,showpacs,floatfix]{revtex4}

\usepackage{amssymb}
\usepackage{graphicx}
\usepackage{dcolumn}
\usepackage{bm}

\begin{document}

\title{Transport through quantum dots in mesoscopic circuits}
\author{P. S. Cornaglia}
\author{C. A. Balseiro}
\affiliation{Instituto Balseiro and Centro At\'omico Bariloche, Comisi\'on Nacional de Energ\'{\i}a At\'omica, 8400 San Carlos de Bariloche, R\'{\i}o Negro, Argentina.}

\begin{abstract}
We study the transport through a quantum dot, in the Kondo Coulomb
blockade valley, embedded in a mesoscopic device with finite wires. The
quantization of states in the circuit that hosts the quantum dot gives rise
to finite size effects. These effects make the conductance sensitive to the
ratio of the Kondo screening length to the wires length and provide a way of
measuring the Kondo cloud. We present results obtained with the numerical
renormalization group for a wide range of physically accessible parameters.
\end{abstract}
\maketitle
Since the pioneer work of Goldhaber-Gordon and coworkers reporting the observation of Kondo effect in a single quantum dot (QD) \cite{GG}, many different circuit and dot configurations have been designed and studied  \cite{QDE}. In a single electron transistor or QD built on a semiconductor heterolayer, the most relevant parameters can be controlled by applying voltages. The possibility of their continuous variation allows to investigate different regimes with different number of electrons localized in the dot \cite{QDT}. States with a well defined number of electrons tend to be stabilized by the Coulomb interaction, a phenomenon known as Coulomb blockade. When an odd number of electrons is stable in the dot and the total spin is $S=1/2$, the coupling with the leads gives rise to the usual Kondo effect. The Kondo effect is the magnetic screening of the dot spin by the electrons of the host \cite{kondo}.
The screening occurs by the formation of a spin singlet involving the dot
spin and the host electron's spins. This screening is nearly fully developed
below a characteristic temperature $T_{K}$ known as the Kondo temperature.
The size of the screening cloud, the spatial extension of the singlet wave
function, is the Kondo screening length $\xi _{K}\approx \hbar v_{F}/T_{K}$
where $v_{F}$ is the Fermi velocity. For a typical QD the Kondo temperature $%
T_{K}$ is of the order of magnitude of one degree and the Kondo screening
length can be up to one micron. In these circuits, the conductance through
the QD and their temperature dependence gives information on the occurrence
of the Kondo effect and ultimately on the development of the Kondo cloud.
The complete development of the Kondo effect is reflected by an ideal
conductance $2e^{2}/h$ as observed in some low temperature measurements \cite
{QDE}.

The smallness of $T_{K}$ makes it possible to alter the Kondo ground state by finite size effects. It was shown that whenever the characteristic size of the system is reduced and the mean level spacing $\Delta $ becomes of the order or larger than $T_{K}$, finite size effects become important \cite{finite,nos1_termo,nos2_din,affleck1,affleck}. Then, in any nanoscopic system with a QD coupled to one dimensional leads or wires of the order of one micron of length, the Kondo effect may be subject to size effects. Based precisely on these effects, Simon and Affleck made two proposals to measure the Kondo screening length \cite{affleck1,affleck}. The first one concerns a closed loop with a QD. The persistent current induced by a magnetic flux threading the ring is sensitive to the screening length and is reduced when the circumference of the ring is smaller than $\xi _{K}$.  The second proposal, which is also the subject of the present work, considers a QD coupled to mesoscopic leads. 
\begin{figure}[tbp]
\includegraphics[width=8.0cm,clip=true]{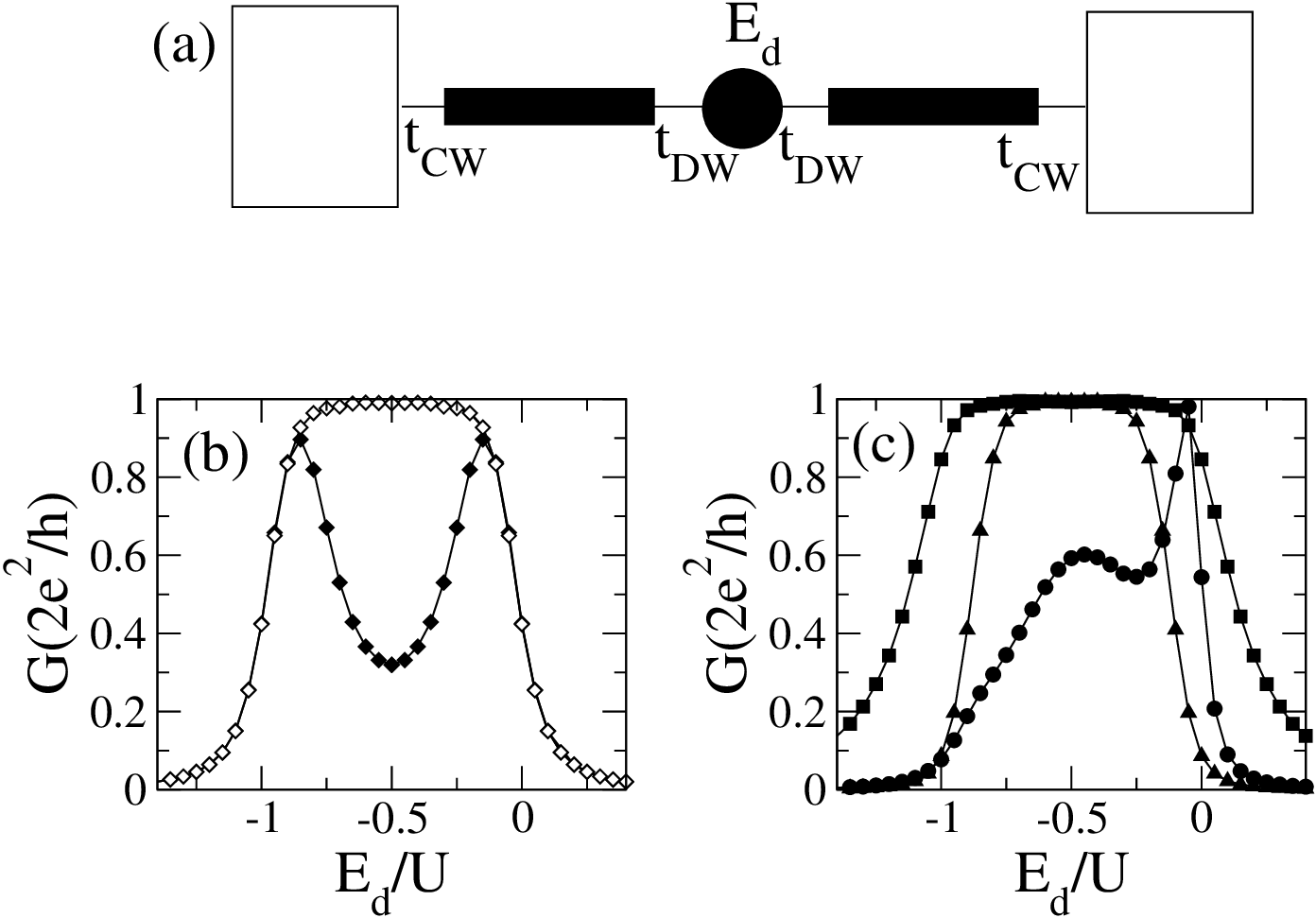}
\caption{(a) Mesoscopic circuit with an embedded quantum dot. (b) Conductance
as a function of the QD level position for a system with infinite wires ($t_{DW}=0.2\,t_0$). $T=0$ (open diamonds) $T = 10\,T_{K}$ (filled diamonds). (c) Conductance as a function of the QD level position for a system with finite wires ($\Delta \simeq 10\,T_{K}^{0}$ and $t_{CW} =0.6\,t_0$) and different values of $\epsilon _{W}$: at-resonance (squares), off-resonance (triangles), and intermediate case $\epsilon _{W}=0.4\,\Delta$ (circles).}
\label{modelo}
\end{figure}

In what follows we study a QD attached to quantum wires as schematically shown in Fig. \ref{modelo}(a). By a quantum wire we mean a narrow wire with a small number of channels. The wires are weakly coupled at one end to the QD and at the other to three dimensional macroscopic contacts that act as a reservoir. The Hamiltonian of the system is then given by 
\begin{equation}
H=H_{D}+H_{W}+H_{C}+H_{DW}+H_{CW},
\end{equation}
where the first three terms corresponding to the dot, wire and contact
Hamiltonians are 
\begin{equation}
H_{D}=\sum_{\sigma }E_{d}d_{\sigma }^{\dagger }d_{\sigma }+Ud_{\uparrow
}^{\dagger }d_{\uparrow }d_{\downarrow }^{\dagger }d_{\downarrow },
\end{equation}
where \ $d_{\sigma }^{\dagger }$ creates an electron with spin $\sigma $ and
energy $E_{d}$ in the quantized state of the QD and $U$ is the Coulomb
repulsion for electrons in this state,
\begin{equation}
H_{W} =\sum_{\eta ,\sigma ,n}^N\varepsilon _{W}c_{\eta n\sigma
}^{\dagger }c_{\eta n\sigma }  
-t_{0}\sum_{\eta ,\sigma ,n}^{N-1}\left( c_{\eta n\sigma }^{\dagger
}c_{\eta (n+1)\sigma }+h.c.\right),
\end{equation}
with $\eta =R,L$ denoting the right and left wires described by the
conventional one dimensional tight binding model. For simplicity, as in Ref. \onlinecite{affleck}, the contact Hamiltonian $H_C$ describes two linear chains with a hopping matrix element $t_{0}$.  The last two terms of the Hamiltonian describe the coupling of these three components: 
$H_{DW}=-t_{DW}\sum_{\eta ,\sigma }\left(d_{\sigma }^{\dagger }c_{\eta 1\sigma
}+h.c.\right)$ and $H_{CW}=-t_{CW}\sum_{\eta ,\sigma }\left(c_{\eta N\sigma }^{\dagger }a_{\eta 1 \sigma }+h.c.\right)$, where $a_{\eta 1 \sigma }$ creates an electron in the first site of the contact $\eta$. 
\begin{figure}[tbp]
\includegraphics[width=8.0cm,clip=true]{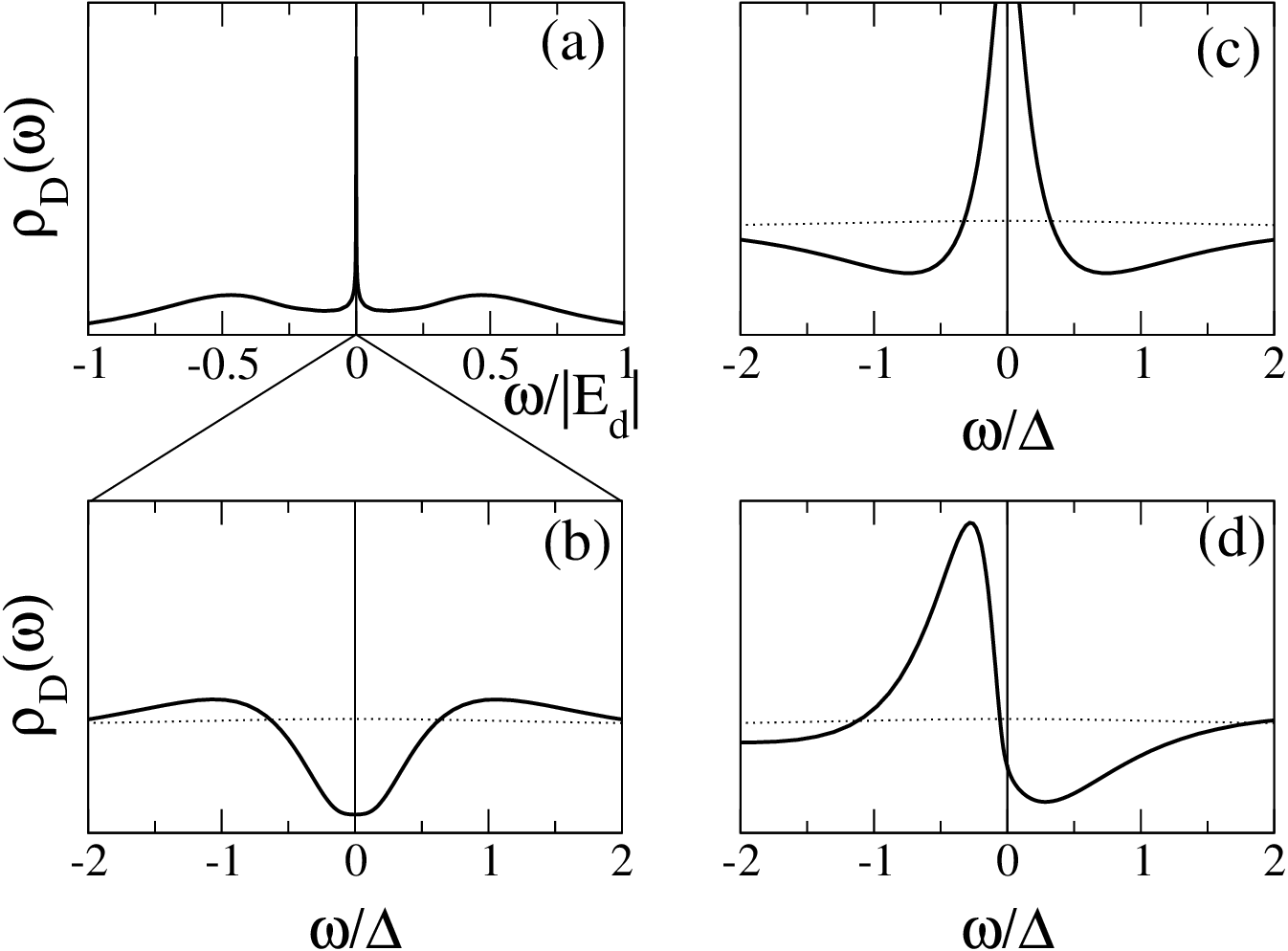}
\caption{QD spectral density (a) and a detail of the Kondo resonance for a system with finite wires: for the at-resonance case (b), off-resonance case (c) and intermediate case (d). The parameters are $E_d=-0.5\,U$, $t_{DW}=0.2\,t_0$, $\Delta = 10\,T_K^0$, and $t_{CW}=0.6\,t_0$. The dotted line in (b), (c), and (d), is the spectral density for infinite wires.}
\label{densesp}
\end{figure}
The parameters $E_d$ and $U$ can be estimated and related to a gate voltage using a capacitance model for the QD \cite{capacitance}.

If $t_{DW}=t_{CW}=0$ the wires are isolated and their energy spectrum corresponds to a set of $N$ states. Around the Fermi energy these states are separated by a characteristic energy $\Delta\simeq \hbar v_{F}\pi /L\simeq 4t_{0}/N$ where $L=aN$ is the wire length and $a$ the lattice constant. When the wires are connected to the contacts with
a non zero $t_{CW}$, the wire states become resonances of width $\gamma$. The local density of states at the wire end $\rho
_{W}(\omega )$, consists of a collection of resonant states characterized by
the two energy scales $\Delta $ and $\gamma $ \cite{nos2_din}. This structure of the system
that hosts the QD may drastically change the Kondo screening and
consequently the linear conductance of the circuit.  Using the Keldysh
formalism \cite{Meir} the conductance can be put as: 
\begin{equation}\label{conduct}
G(T)=\frac{2e^{2}}{h}2\pi \int d\omega (-\frac{\partial f(\omega )}{\partial
\omega })\Gamma (\omega )\rho_{D}(\omega )  \label{cond}
\end{equation}
where $f(\omega )$ is the Fermi function, $\Gamma (\omega )=\pi
(t_{DW})^{2}\rho _{W}(\omega )$ and $\rho _{D}(\omega )\mathcal{\ }$is the
QD spectral density \cite{kondo}. This simple formula gives the conductance of the central part of the circuit: QD plus wires. It can be obtained calculating the
current through the CW-links \cite{Meir} and expressing it in terms of the QD spectral density. The spectral densities are calculated using the numerical renormalization group (NRG) technique \cite{nrg1,nrg2}. The linear conductance as function of $E_{d}$ is shown in Fig. \ref{modelo}(b) for $t_{CW}=t_{0}$ corresponding to long wires ($L\rightarrow \infty $) without constrictions. At high temperatures two Coulomb peaks and the central Coulomb blockade valley are clearly observed.
The Coulomb peaks are due to $E_{d}$ or $E_{d}+U$ being aligned with the
Fermi energy. The central valley, away from the Coulomb blockade peaks, has one electron localized at the dot and corresponds to the Kondo regime. As the temperature is lowered, the conductance at the central Kondo valley increases indicating the occurrence of the Kondo screening. In this case $\rho _{W}(\omega )$ is constant around the Fermi level and $\rho _{D}(\omega )$ develops a Kondo resonance as shown in Fig. \ref{densesp}(a). At zero temperature at the center of the Kondo valley 
$2\pi \Gamma (E_{F})\rho _{D}(E_{F})=1$, a situation known as the unitary
limit, and the conductance is simply $2e^{2}/h$.  For finite wires with a
constriction ($t_{CW}<t_{0}$) that generates a fine structure with peaks
separated by $\Delta $ in $\rho _{W}(\omega )$, the low temperature
conductance shows finite size effects in the whole range of $E_{d}$ as shown
in Fig. \ref{modelo}(c). At these low temperatures, the conductance becomes
sensitive to the relative position of the Fermi energy $E_{F}$ and the structure of $\rho _{W}(\omega )$. We distinguish three cases: {\it i}) the
Fermi level lying at a wire resonant state, a maximum in $\rho _{W}(\omega )
$ (the at-resonance case), {\it ii}) exactly between two resonances, a minimum in $\rho _{W}(\omega )$ (the off-resonance case) and {\it iii}) intermediate situations. As shown in \ref{modelo}(c) for the at-resonance and off-resonance cases, a conductance of $2e^2/h$ is obtained at the Coulomb blockade valley, while for intermediate situations strong anomalies are obtained. These results correspond to the low temperature limit, for $T>\Delta$ the structure of  $\rho _{W}(\omega )$ becomes unimportant and all three curves of Fig. \ref{modelo}(c) collapse into a single one that reproduces the high temperature behavior of Fig. \ref{modelo}(b).

The behavior of the transmission close to the Coulomb peaks can be understood in terms of a simple single particle resonant state lying above (for $E_{d}\gtrsim 0$) or below (for $E_{d}\lesssim -U$) the Fermi energy. If $E_{d}\simeq -U/2$, the center of the Coulomb blockade valley, the behavior is dominated by the Kondo physics. Let us now
concentrate in the Kondo regime ($E_{d}<0$ and $E_{d}+U>0$) and define a
reference Kondo temperature $T_{K}^{0}$ \cite{nrg1} of the system with infinitely long
quantum wires ($L\rightarrow \infty $). If a system with finite wires is
such that $T_{K}^{0}\gg \Delta $, on the scale of the characteristic Kondo
energy the host local density of states can be averaged to its mean value
and the finite size effects are washed out for any temperatures $T\gtrsim
T_{K}^{0}$. This means that as the temperature is lowered and the Kondo
screening starts to develop, the finite size effects are unimportant.
Conversely, if the system were such that $T_{K}^{0}\ll \Delta $, we expect
strong finite size effects at any temperature $T\lesssim \Delta $, i.e. even
before the Kondo effect of the reference system starts to develop.
\begin{figure}[tbp]
\includegraphics[width=8.0cm,clip=true]{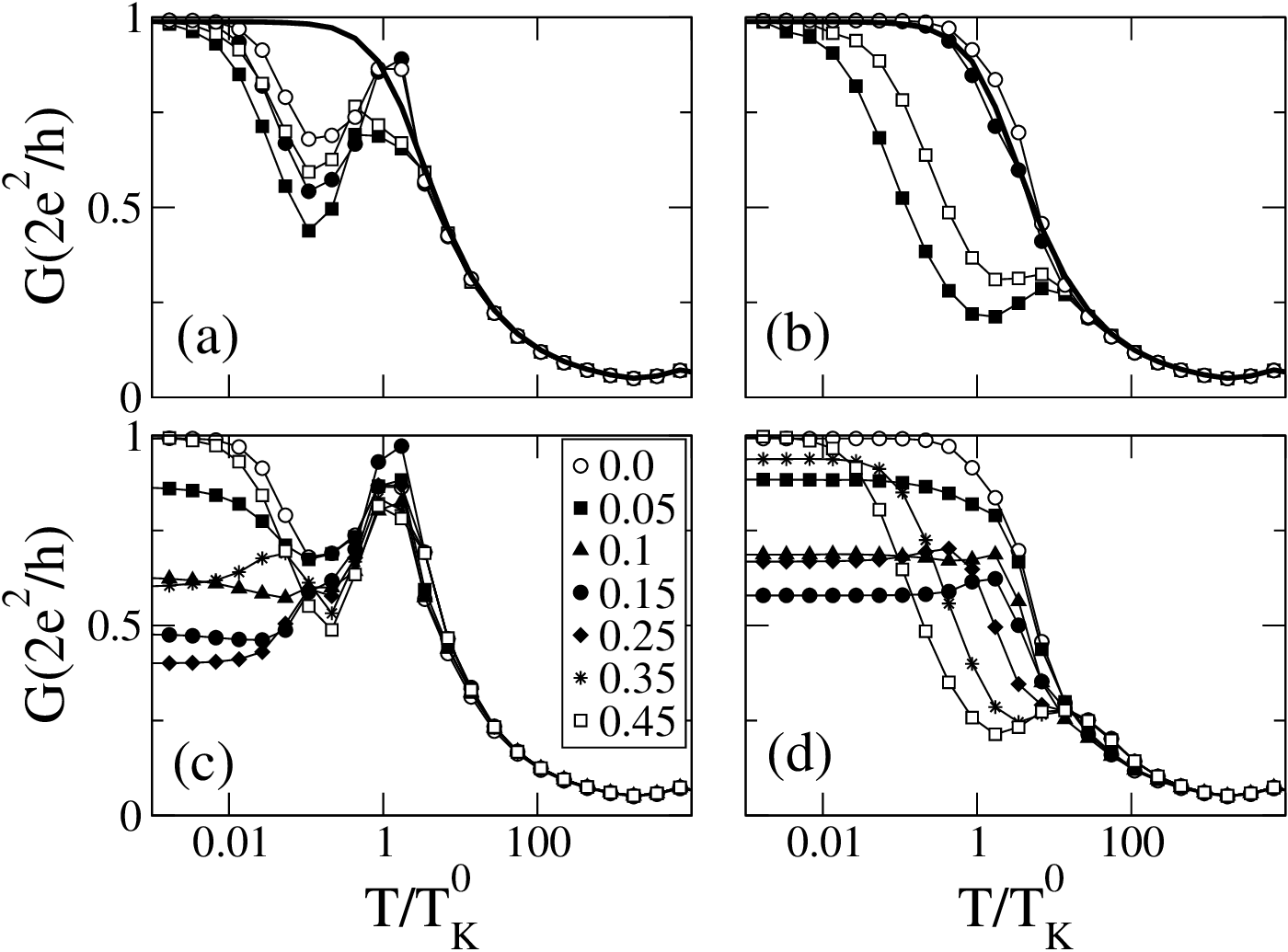}
\caption{Conductance as a function of temperature for a $\Delta = T_K^0/3$ system (a) and (c), and a $\Delta = 10\,T_K^0$ system (b) and (d). (a) and (b) At-resonance (circles) and off-resonance (squares) situations with $t_{CW}=0.5\,t_0$ (filled symbols), and $t_{CW}= 0.6\,t_0$ (open symbols). (c) and (d)  Conductance for different values of $\epsilon_W$ and $t_{CW}=0.5\,t_0$. The other parameters are $E_d=-0.5\,U$ and $t_{DW}=0.2\,t_0$.}
\label{condT}
\end{figure}

In a circuit built on a semiconductor heterolayer, the position of $E_{F}$ relative to the wire structure as well as the coupling to the reservoirs ($t_{CW}$) can be varied applying gate voltages \cite{QDE}. In Fig. \ref{condT} the
conductance as a function of temperature for different values of the parameters is shown. In what follows we analyze these results according to the position of the Fermi level relative to the structure of the wire density of states $\rho
_{W}(\omega )$.

\textit{i) At-resonance case: }For $t_{CW}=0$ the central QD with the attached
wires form an isolated system and the QD spectral density is given by a
collection of delta-functions. For this case of a small isolated system,
which has been studied in some detail \cite{nos2_din}, the QD spectral density $\rho
_{D}(\omega )$ has two delta-functions with relatively strong weight around
the Fermi energy. One is just above and the other just below $E_{F}$. As
the system is coupled to the macroscopic contacts ($t_{CW}\neq 0$) the delta
functions in $\rho _{D}(\omega )$ acquire a finite width. However, for weak
coupling the double structure around the Fermi energy is clearly observed
and the Kondo resonance has a minimum at\ $E_{F}$. In Fig. \ref{densesp}(b) a low frequency detail of the QD spectral densities for the at-resonance case is
shown. As the temperature increases, the whole structure is washed out. The conductance obtained from Eq. (\ref{conduct}) for a relatively long quantum wire
with $T_{K}^{0}=3\Delta $ is shown in Fig \ref{condT}(a) for different values of the wire-contact coupling strength $t_{CW}/t_{0}$. For an ideal coupling ($t_{CW} =t_)$) the conductance increases as $T$ decreases to reach the
value $2e^2/h$.  For $t_{CW}<t_0$, as $T$ is lowered and approaches the energy scale $\Delta $, the structure in $\rho _{D}(\omega )$ becomes relevant and the conductance departs from the $t_{CW} =t_0$ case. As $T\rightarrow 0$, the ideal value is recovered generating a minimum in the conductance. In the low
temperature regime, the QD acts as a perfect link between the right and left
wires creating a single wire of length $2L$. The at-resonance condition
implies that the Fermi level is aligned with a wire state giving an ideal
conductance. For short wires with  $T_{K}^{0}<\Delta $, the screening develops for temperatures $T\sim\Delta$ \cite{nos1_termo} and the conductance is not very sensitive to the confinement effects [Fig. \ref{condT}(b)].
\begin{figure}[tbp]
\includegraphics[width=8.0cm,clip=true]{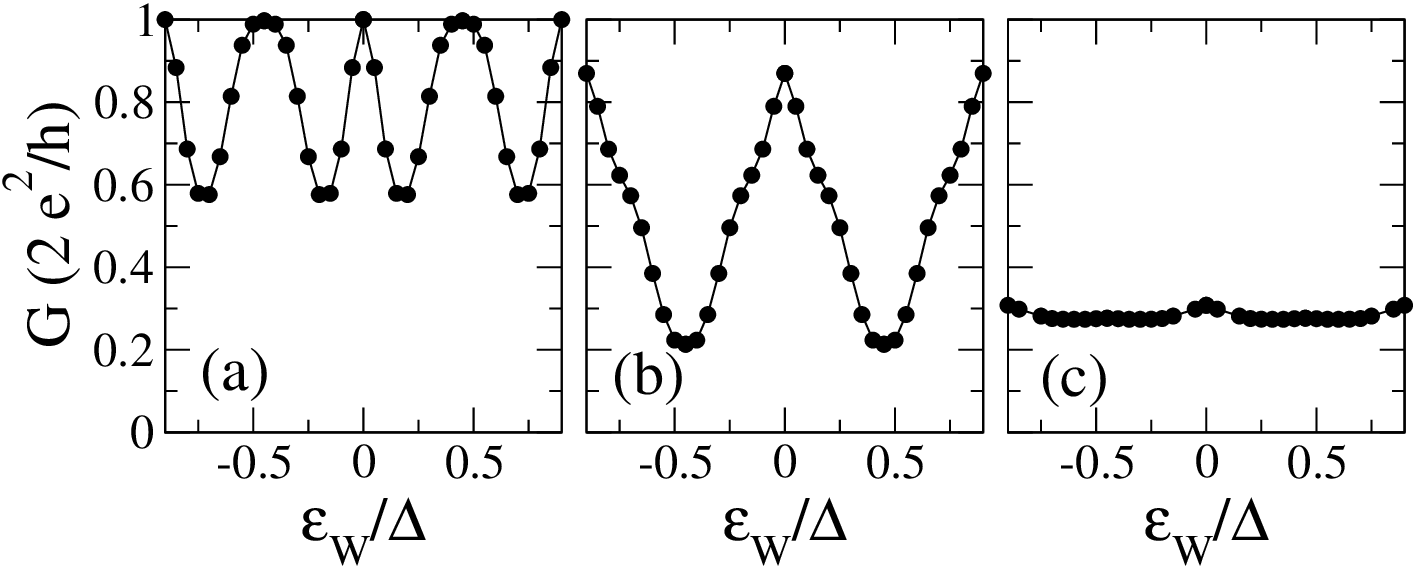}
\caption{ Conductance as a function of $\epsilon _{W}$ for a $\Delta
=10\,T_{K}^{0}$ system at $T=0$ (a), $T\simeq T_{K}^{0}$ (b), and $T=\Delta$ (c).  $\varepsilon_W/\Delta=0$, $\pm0.5$ correspond to the at-resonance and off-resonance cases respectively.}
\label{G_ew}
\end{figure}

\textit{ii) Off-resonance case: }. For $T_{K}^{0}>\Delta $ and high temperatures the fine structure if $\rho_W(\omega)$ is not important
and the conductance for the at-resonance and off-resonance cases behaves in the same way. For low temperatures, again the QD acts as a perfect link, the resulting effective wire of length $2L$ has resonant states separated by $\Delta /2$, rather than by $\Delta$, and one of them is aligned with $E_{F}$. Although the at-resonance an off-resonance spectral densities are quite different [Figs. \ref{densesp}(b) and \ref{densesp}(c)], the
temperature dependence of the conductance is similar [Fig. \ref{condT}(a)],
in fact the product $\Gamma (\omega )\rho _{D}(\omega )$ has qualitatively
the same structure in both cases giving a conductance that does not clearly
distinguish the two situations. For short wires, as the temperature increases, the conductance rapidly diminishes from its low temperature value $G\sim 2e^2/h$ [Fig. \ref{condT}(b)].

\textit{iii) Intermediate case:} This situation where the Fermi
level lies at an arbitrary position with respect to wire states generates
a quite different behavior at low temperatures. Again for $T>\Delta $  the
quantization effects are irrelevant and the conductance is not distinguished
from that of the previous cases. At low temperatures the conductance, never reaches the value $2e^2/h$ [see full symbols in Figs. \ref{condT}(c) and \ref{condT}(d)]. Even if the QD were behaving as a perfect link, the resulting $2L$ wire states would not be aligned with $E_{F}$.  

\begin{figure}[tbp]
\includegraphics[width=8.0cm,clip=true]{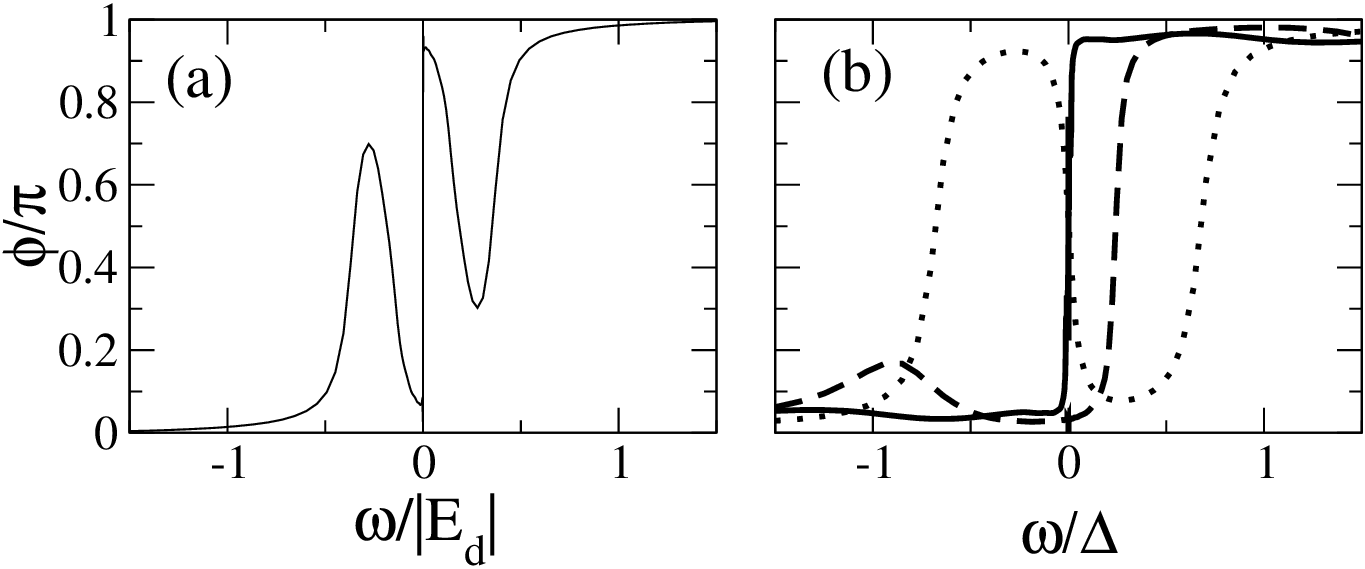}
\caption{(a) Zero temperature transmission phase shift for a $\Delta =10\,T_{K}^{0}$ system. (b) Low energy detail for the at-resonance case (dotted style), off-resonance case (solid style) and an intermediate case (dashed style).} 
\label{phaseS} 
\end{figure}
The conductance as a function of $\varepsilon _{W}$ for short ($T_{K}^{0}<\Delta $) wires is shown in Fig. \ref{G_ew} for different temperatures. As discussed above, for low temperatures both the at-resonance and off-resonance cases give a conductance close to its ideal value $2e^2/h$, as shown in Fig. \ref{G_ew}(a). In agreement with the results of Ref. \onlinecite{affleck}, for intermediate temperatures ($T\sim T_K^0$), the off-resonance case gives a small conductance [see Fig. \ref{G_ew}(b)]. This change of behavior looks like a change in the periodicity of $G$ vs $\varepsilon_W$. Finally at high enough temperatures the amplitude of the oscillations goes to zero [see Fig. \ref{G_ew}(c)]. For long wires, as can be deduced from Fig. \ref{condT}(c), the intermediate regime is never observed i.e. the system does not show a doubling of periodicity in $G$ vs $\varepsilon_W$. In quantum wires with one channel, the condition $T_{K}^{0}\gtrless \Delta $ is equivalent to $L\gtrless \xi _{K}^{0}$ and these results of $G(T)$ vs $\varepsilon _{W}$ could be used to measure the Kondo screening length.

Finally we analyze the finite size effects on the phase shift introduced by
the QD \cite{oreg}. We define an energy dependent transmission  as $t_{\sigma }(\omega )=\Gamma (\omega )G_{D\sigma }(\omega )$
where $G_{D\sigma }(\omega )$ is the retarded Green's function for a spin-$%
\sigma $ electron in the QD level. The conductance  as given by equation (%
\ref{cond}) is the thermal average of $-1/\pi \sum_{\sigma }Im[t_{\sigma }(\omega )]$. The phase of the transmission  for a spin-$\sigma $
electron, $\phi _{\sigma }(\omega )=\arg [t_{\sigma }(\omega )]$, is shown
in Fig. \ref{phaseS} for a system with finite wires. On a large energy scale, the low temperature behavior of $\phi _{\sigma }(\omega )$ qualitatively reproduces the results of Gerland et al.\cite{Costi}, it has a maximum (minimum) for $\omega \approx E_{d}/2$ [$\omega \approx (E_{d}+U)/2$] and a large phase lapse for $\omega \approx 0$. This large phase lapse at zero frequency shows novel features due to the confinement effects. A zoom of the low frequency details  in Fig. \ref{phaseS}(b), shows the behavior of $\phi _{\sigma }(\omega )$ for $\omega \lesssim \Delta $ with a superstructure consistent with that of the Kondo peak. These effects in the transmission phase shift would determine the current in Aharonov-Bohm interferometers with two arms and an embedded QD.  

A logical step in the development of molecular electronics is the connection of quantum wires to single electron transistors to be used as building blocks. The reduction of the quantum wires dimensions down to the micron lengthscale may change the system properties. For such devices to be useful, a detailed knowledge of the behavior of the transmission phase shift and the conductance trough this system is needed. In this paper we have shown how these properties change when the QD is connected to finite quantum wires. In particular, the behavior of the system is very sensitive to the length of the quantum wires and to the position of the Fermi level relative to the structure of the local density of states.  The confinement introduces anomalous features in the temperature dependence of the conductance and the energy dependence of the phase shift. 
For single channel wires in the Kondo regime the response of the system depends on whether the Kondo screening lenght is shorter or larger than the quantum wire length. Finite size effects are not only relevant in the Kondo regime, the Coulomb blockade peaks of the conductance are also strongly affected as shown in Fig. \ref{modelo}.
 
This work was partially supported by the CONICET and ANPCYT,
Grants N. 02151 and 99 3-6343.

\end{document}